\documentclass[12pt, draftclsnofoot, onecolumn]{IEEEtran}

\usepackage{amsthm}
\usepackage[cmex10]{amsmath}
\usepackage{graphicx}
\usepackage{color}
\usepackage[tight,footnotesize]{subfigure}
\usepackage{amsfonts}
\usepackage{algorithmic}
\usepackage{algorithm}
\usepackage{graphicx}
\usepackage{subfigure}
\usepackage{bm}
\usepackage{times,color,amssymb,epsfig,psfrag,stfloats,enumerate}
\usepackage{cite}

\newtheorem{theorem}{Theorem}

\newtheorem{lemma}{Lemma}

\ifCLASSINFOpdf
\else
\fi
\begin{document}
%
\title{Policy Optimization for Content Push via Energy Harvesting Small Cells in Heterogeneous Networks}

\author{Jie~Gong~\IEEEmembership{Member,~IEEE}, Sheng~Zhou~\IEEEmembership{Member,~IEEE}, Zhenyu~Zhou~\IEEEmembership{Member,~IEEE}, Zhisheng Niu~\IEEEmembership{Fellow,~IEEE}

\thanks{Jie Gong is with School of Data and Computer Science, Sun Yat-sen University, Guangzhou 510006, P.~R.~China. Email: gongj26@mail.sysu.edu.cn.}
\thanks{Sheng Zhou and Zhisheng Niu are with Tsinghua National Laboratory for Information Science and Technology, Department of Electronic Engineering, Tsinghua University, Beijing 100084, China. Emails: \{sheng.zhou, niuzhs\}@tsinghua.edu.cn.}
\thanks{Zhenyu Zhou is with State Key Laboratory of Alternate Electrical Power System with Renewable Energy Sources, School of Electrical and Electronic Engineering, North China Electric Power University, Beijing 102206, China. Email: zhenyu\_zhou@ncepu.edu.cn}
\thanks{This work is sponsored in part by the Fundamental Research Funds for the Central Universities, the National Basic Research Program of China (973 Program: No.2012CB316001), the Nature Science Foundation of China (61571265, 61321061, 61601180, 61601181, and 61461136004) and Hitachi R\&D Headquarter.}
}

\maketitle

\begin{abstract}
Motivated by the rapid development of energy harvesting technology and content-aware communication in access networks, this paper considers the push mechanism design in small-cell base stations (SBSs) powered by renewable energy. A user request can be satisfied by either push or unicast from the SBS. If the SBS cannot handle the request, the user is blocked by the SBS and is served by the macro-cell BS (MBS) instead, which typically consumes more energy. We aim to minimize the ratio of user requests blocked by the SBS. With finite battery capacity, Markov decision process based problem is formulated, and the optimal policy is found by dynamic programming (DP). Two threshold-based policies are proposed: the push-only threshold-based (POTB) policy and the energy-efficient threshold-based (EETB) policy, and the closed-form blocking probabilities with infinite battery capacity are derived. Numerical results show that the proposed policies outperform the conventional non-push policy if the content popularity changes slowly or the content request generating rate is high, and can achieve the performance of the greedy optimal threshold-based (GOTB) policy. In addition, the performance gap between the threshold-based policies and the DP optimal policy is small when the energy arrival rate is low or the request generating rate is high.
\end{abstract}

\IEEEpeerreviewmaketitle

\section{Introduction}
With the rapid growth of the wireless multimedia traffic, the enormous energy consumption of wireless communication systems has become one of the major concerns in the future 5G networks. Energy harvesting (EH) technology \cite{gunduz2014designing, sharma2010optimal, yang2012optimal, ozel2011transmission, gong2013optimal, tao2013on, gong2014base}, which utilizes the renewable energy from natural sources such as solar, wind, kinetic activities and so on, can greatly reduce the power consumption from the conventional grid power supply, i.e., power grid, and hence reduce the $\mathrm{CO}_2$ emissions. It has been considered as one of the candidate technologies to achieve green communications. On the other hand, the heterogeneous wireless network, where the small-cell base stations (SBSs) with low transmit power take in charge of a small area to enhance the overall performance, is considered as a possible future network architecture. The SBSs can be powered by renewable energy to reduce deployment and operation costs \cite{piro2013hetnets, Yuyi2015energy}. During the periods of energy deficiency, the users can instead be served by the macro-cell base stations (MBSs) with grid power supply. However, due to the randomness of the renewable energy arrival and the limitation on the battery capacity, energy shortage or waste will occur when the traffic pattern mismatches the energy harvesting process in either spatial or time domain. Content caching and push \cite{podnar2002distributed, nico2002using, wang2014push, golrezaei2013femtocaching, Shanmugam2013femtocaching, Poul2014multicast, Blasco2014learning, Bastug2014cache, wang2011on, bastug2014living, sadr2013anticipatory, Maria2015joint} at access points and end users can effectively reduce the duplicated content transmissions, and hence, reduce the core network traffic load as well as the energy consumption. As a result, it is promising to deal with the mismatch between energy and traffic via content caching and push in renewable energy powered small cells.

In the literature, there have been extensive studies on how to match harvested energy with traffic profile. The stability condition and the average delay of the queuing system with both energy queue and data queue are analyzed in \cite{sharma2010optimal}. The work in \cite{yang2012optimal} studies the optimal power allocation for transmission time minimization with random packet arrival under additive white Gaussian noise (AWGN) channel. The power allocation problem is further extended to the fading channel case in \cite{ozel2011transmission, gong2013optimal}. The long-term spatial and temporal variations of mobile traffic are considered in \cite{tao2013on, gong2014base}, and the renewable energy is adapted through resource allocation and base station (BS) sleeping, respectively. {The small BS on/off scheduling in heterogeneous networks is also studied in \cite{lee2016online} to minimize energy consumption and transmission delay. There are other ways to adjust the energy profile between nodes, for example, devices can trade the harvested energy with one another based on the service requirement \cite{xiao2015dynamic}, and the secondary user in cognitive radio networks can harvest the RF energy from the primary user and use it to transmit in idle times \cite{hoang2014opportunistic}.} The existing studies mainly focus on the algorithm design based on the traffic intensity, i.e., how many bits or packets need to be transmitted. However, the important content information carried inside the bit stream has not been well explored.

In this paper, we try to further improve the performance by exploring the content information. Our work can be viewed as part of the recently proposed GreenDelivery framework \cite{Zhou2014greendelivery}, which adopts content caching and push in EH powered SBSs to enable efficient content delivery. Statistic data shows that people's interests to the contents are not equal \cite{cha2007tube}. Wireless multicast \cite{niu2008new, liu2013utility} holds the promise of achieving huge energy efficiency gain via broadcasting commonly interested multimedia contents to multiple users simultaneously. Hence, the content push mechanism \cite{podnar2002distributed} is developed based on multicast to make full use of the content popularity features. A learning algorithm for adapting the proactive push to the dynamic client demands is proposed in \cite{nico2002using}. And the capacity gain by push in the integrated broadcast and communication network is analyzed in \cite{wang2014push}. In reality, proactive push is supported by the development of last-mile wireless access hardwares and mobile devices. To reduce the core network overhead, contents are suggested to be cached at the SBSs \cite{golrezaei2013femtocaching, Shanmugam2013femtocaching, Poul2014multicast, Bastug2014cache, Blasco2014learning} or relay nodes \cite{wang2011on}, with proactive caching schemes \cite{bastug2014living, sadr2013anticipatory, Maria2015joint}. The work in \cite{zhou2015stochastic} uses multicast to satisfy users' common requests in the cache-enabled heterogeneous cellular networks and formulates the problem by Markov decision process (MDP) \cite{bertsekas2005dynamic}. However, pushing contents prior to the actual requests is not considered. In addition, the dynamic energy arrival is not considered either. Recently, EH based SBSs are used to cache contents \cite{sharma2013greencache} for improving deployment flexibility and reducing energy consumption. In \cite{kumar2015tradeoff}, an online power control scheme is developed for EH based SBS with wireless backhaul and local cache. On the other hand, with the rapid improvement of data storage capacity, user devices are capable of storing large amount of data. With the available contents at the SBSs and the large user storage capacity, proactive push by EH powered SBSs in heterogeneous cellular networks is considered to be practical and effective.

Combining EH with content push provides potentials to improve the efficiency of the renewable energy. Specifically, the contents are pushed to users earlier than the actual demands when the harvested energy is sufficient. In reward, the energy waste due to the battery overflow can be avoided as the harvested energy can be used effectively and timely. Later, when the energy state becomes poor, the content requests can be satisfied by users' local storage. It can be regarded as \emph{transferring} the harvested energy along with the timeline to the \emph{future} to match the content requests. However, the push mechanism design in EH powered SBSs is still an open problem, which is not trivial since it needs to jointly consider energy state, traffic load as well as content popularity. The advantage of content push is that it reduces the duplicated transmissions of the same content. But at the same time, pushing a content typically consumes more energy than unicast as it needs to guarantee the data rate of the worst-channel user. Hence, there is a tradeoff between the energy saving due to the reduction of the duplicated content transmissions and the energy consumption of content push. We try to explore the tradeoff and design efficient push mechanisms. Notice that if the user request cannot be satisfied in the small-cell and has to be handled by the MBS, we say that it is blocked by the SBS. The objective is to minimize the blocking probability in the SBS. The preliminary optimal policy design has been presented in \cite{gong2015proactive}. This paper extensively considers the design of policies with low-complexity. The contributions are as follows:

\begin{itemize}
 \item[$\bullet$] With finite battery capacity, we formulate the problem using MDP tool by discretizing the system state (including the battery state, the content request and {the set of contents which have been pushed}) into a finite set, and find the optimal stationary policy via dynamic programming (DP).
 \item[$\bullet$] We propose two threshold-based policies: The push-only threshold-based (POTB) policy and the energy-efficient threshold-based (EETB) policy. With infinite battery capacity, we derive the closed-form blocking probability which equals to the result under average power constraint. In finite battery capacity, the threshold-based policies can be analyzed via the finite state Markov chain (FSMC) model.
 \item[$\bullet$] The performances of the proposed policies are evaluated via numerical simulations compared with the conventional non-push policy, the greedy optimal threshold-based (GOTB) policy and the DP optimal policy. The sensitivity of the proposed threshold-based policies to the battery capacity is also studied.
\end{itemize}

The rest of the paper is organized as follows. Section \ref{model} presents the system model and the problem formulation. Section \ref{sec:MDP} provides the MDP analysis and the policy iteration algorithm to find the optimal policy. The threshold-based polices are proposed and analyzed in Section \ref{sec:thrpolicy}. Numerical results are provided in Section \ref{simulation} for performance evaluation. Finally, Section \ref{concl} concludes the paper.

\begin{figure}
\centering
\includegraphics[width=3.4in]{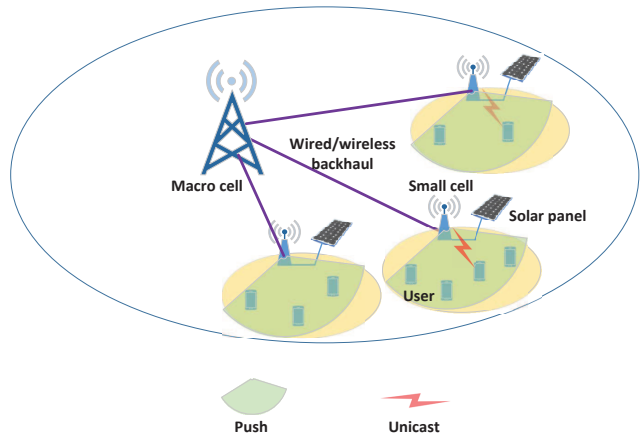}
\caption{Two-tier heterogeneous cellular network. {SBSs can fetch contents via backhaul link and store in their cache. The cached contents can be either pushed to all users or unicasted to one of them.}} \label{fig:system}
\end{figure}

\section{System Model and Problem Formulation} \label{model}
{We consider a two-tier heterogeneous cellular network composed of a MBS and multiple SBSs as shown in Fig.~\ref{fig:system}. Each SBS serves the users in its coverage and treats the data transmissions of other SBSs as background interference. Since the SBSs are densely deployed, the sum interference is considered the same for each user. Hence, we only focus on the policy design of a single SBS and all the SBSs can work in the same way.} The MBS is powered by the power grid and the SBS is powered by renewable energy solely. The harvested energy is stored in a battery with finite capacity. The SBS is connected with the MBS via a wired/wireless backhaul link to fetch contents and store them in its cache to reduce the congestion of core networks. We assume in this paper the time and energy consumption to fetch contents is negligible, and thus only focus on how to push contents from the SBS to users. 

\begin{figure}
\centering
\includegraphics[width=3.4in]{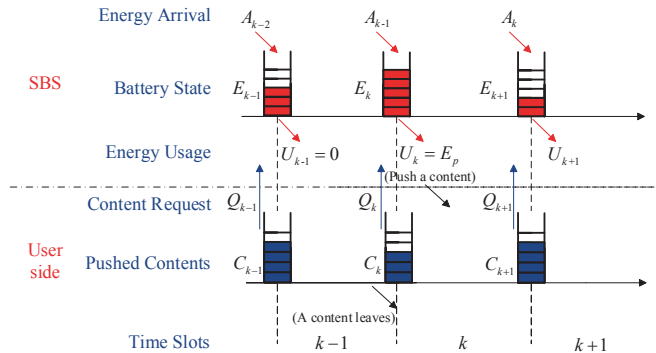}
\caption{{Timeline of the slotted system and examples of actions. At the beginning of slot $k-1$, the SBS takes sleep action ($U_{k-1} = 0$). By the end of slot $k-1$, a content leaves the system, and the number of pushed contents $C_k$ is reduced by 1. At the beginning of slot $k$, the SBS takes push action ($U_{k-1} = E_p$), which results in an increase of $C_{k+1}$ by the end of the slot.}} \label{fig:queue}
\end{figure}

Fig.~\ref{fig:queue} illustrates how the dynamic system evolves. Specifically, the system is slotted with slot length $T_p$. In each slot $k$, a content request $Q_k$ is generated from the user side. Based on the request and the battery energy state, the SBS takes an action with energy usage $U_k$, and then harvests a certain amount of energy $A_k$. At the user side, each user receives the contents and preserves a list of pushed contents $\bm{C}_k$. The detailed description of the model is as follows.

\subsection{Content Request and Channel Model}
The content request is assumed to follow the Bernoulli distribution, i.e., there is a content request with probability $p_u\in [0, 1]$ in each slot. And the location of the request follows the uniform distribution in the cell coverage. The reasons for using Bernoulli distribution are as follows. Firstly, due to the limited energy availability, the renewable energy powered SBS is usually deployed with small coverage or in regions of low traffic load, i.e., in most cases, there is no more than one request at one time. If multiple contents are requested simultaneously, they are probably the popular ones which can be satisfied by the users' local storage and will not trigger simultaneous multiple transmissions. As a consequence, our assumption is a good approximation of the real system. {Secondly, the total number of content requests in the macro-cell (summation of independent Bernoulli distributed random variables) follow the binomial distribution, which can be well fitted by a Poisson distribution as the SBSs are densely deployed. Hence, the model for generating content request in macro-cell well meets the typical Independent Reference Model (IRM) provided in \cite{paschos2016wireless}.}

Each content is transmitted in one time slot with fixed target average data rate $r_0$. For each content transmission, the average data rate of a given user can be calculated as
\begin{equation}
r_0 = \mathbb{E}_h\left[W\log_2\left(1+\frac{P_t|h|^2\beta d^{-\alpha}}{\sigma^2+P_I}\right)\right], \label{eq:rate}
\end{equation}
where $W$ is the bandwidth of the SBS, $P_t$ is the transmit power, $h$ is the small-scale fast fading coefficient, $\beta$ and $\alpha$ represent the pathloss constant and the pathloss exponent, respectively, $d$ is the transmission distance, $\sigma^2+P_I$ is the noise plus interference power. Assume the SBSs and the MBS are allocated with orthogonal frequency bands, respectively. As a result, the interference is only caused by the SBSs working in the same frequency band. And the interference power $P_I$ is computed by the summation over all the interfering SBSs. $\mathbb{E}_h$ is the expectation operator with respect to $h$. That is, the average data rate $r_0$ is obtained by averaging the instantaneous channel rates of all fading blocks. In practice, as a content usually contains multiple data packets, each transmission slot contains multiple packet transmission frames, among which the small scale fading $h$ changes. Hence, the data rate averaging over $h$ is a good approximation of the actual content delivery rate. We also assume that the user is of low mobility so that it remains in one cell during transmission. Based on the channel model, the required transmit power $P_t$ can be obtained by solving (\ref{eq:rate}) numerically. This can be done by the bi-section method \cite[Chap.~2.1]{richard1985douglas} for instance. To calculate the rate, the channel state information (CSI) is required. In real networks, the users will get the CSI via standard channel training process. In addition, only a small number of bits will be fed back to the SBS for power control.

Based on the above model, the content request state $Q_k$ can be denoted as the amount of energy required for unicasting a content. As the calculation is averaged over $h$, $Q_k$ is a function of the transmission distance, i.e., $Q_k = E_u(d)$. Notice that $Q_k = 0$ indicates that there is no unicast request.

\subsection{Action and Energy Model of the SBS}
The SBS has three possible actions: sleep, unicast, and push. When the SBS has sufficient energy, it can either unicast a content upon request, or push a popular content to all the users in its coverage. When the battery energy is depleted, the SBS enters the sleep mode and the content request will be handled by the MBS. The action is denoted by $u_k$ and is defined as
\begin{equation}
u_k = \left\{ \begin{array}{ll} 0, & \textrm{sleep} \\
1, & \textrm{unicast~a~content}\\
2. & \textrm{push~a~content}\end{array}\right. \label{eq:uk}
\end{equation}
Notice that when push action is taken, the SBS always pushes the most popular content to the users. This is the most efficient as it minimizes the number of the duplicated data transmissions. The action to be taken is constrained by the available battery energy, i.e., the energy used in slot $k$ satisfies $U_k \le E_k$. In sleep action, we assume the energy consumption is negligible, i.e., $U_k = 0$. In push action, the energy $U_k = E_p$ is used to guarantee that all the users in the cell coverage can successfully receive the content. Then the battery energy state is updated as $E_{k+1} = \min\{B_{\mathrm{max}}, E_k-U_k+A_k\}$, where $B_{\mathrm{max}}$ is the battery capacity, and the harvested energy $A_k$ is assumed ergodic and i.i.d.~with mean $\bar{A}$.

\subsection{Content Popularity Model}
Assume there are in total $N$ contents that the users are interested in, denoted by the list $\bm{C} = (c_1, \ldots, c_N)^T$, where the $i$-th ranked content $c_i$ is requested with probability (or popularity) $f_i$. Statistical researches have shown that the content popularity distribution is well fitted by the Zipf distribution \cite{cha2007tube, golrezaei2013femtocaching}. The popularity of content $c_i$ can be expressed as
\begin{equation}
f_i = \frac{1/i^v}{\sum_{j=1}^N 1/j^v}, \label{eq:pop}
\end{equation}
where $v \ge 0$ is the skew parameter. Obviously, we have $f_1 > \cdots > f_N$.

In addition, as people's interest changes over time, old contents may be outdated and new contents will attract people's interest. We assume in each slot there is an old content replaced by a new one with probability $p_c \in [0, 1]$, and the old one is randomly picked over $\bm{C}$ with equal probability $\frac{1}{N}$. While the popularity changes as follows. If content $c_i$ is replaced by the new content $c'$, the content list becomes $\bm{C}' = (c_1, \ldots, c_{i-1}, c_{i+1}, \ldots, c_N, c')^T$, i.e., the ranks of the contents $c_{i+1}, \ldots, c_N$ increase by one and the new content $c'$ is with rank $N$. Such a content updating model is used for two reasons. Firstly, it reflects the fact that the popularity of most contents increases gradually in real systems \cite{cha2007tube}. Secondly, it simplifies the problem formulation, i.e., we only need to consider the number of popular contents rather than the popularity of each individual content. Specifically, according to the push action and content model, it can be easily verified that the pushed content list is $\bm{C}_k = (c_1, \ldots, c_{C_k})$, where $C_k \le N$ is the number of pushed contents. That is, the users always preserve the most popular contents. The push action is taken on the remained list $\widetilde{\bm{C}}_k = (c_{C_k+1}, \ldots, c_N)$, and the content $c_{C_k+1}$ should be pushed at first. As a result, we can use the scalar $C_k$ instead of the vector $\bm{C}_k$ to indicate the push state.

{In summary, the randomness involved in the system model includes the following. Each content request is generated randomly according to Bernoulli distribution. Which content is requested follows the popularity distribution defined in (\ref{eq:pop}), and its location follows uniform distribution. The random location of the user request indicates the randomness of channel. In addition, the energy arrives randomly, and the content list is updated randomly. As the system state changes among slots, the action should also change accordingly.}

\subsection{Problem Formulation}
When a user requests a content that is not in its storage, and at the same time, the SBS does not unicast the content, the request is blocked by the SBS and needs to be handled by the MBS, which usually causes more energy consumption due to the larger transmission distance. We aim at optimizing the usage of harvested energy to minimize the ratio of content requests blocked by the SBS, which is named as the \emph{blocking probability} at the SBS in the rest of the paper. Mathematically, the objective is expressed as
\begin{equation}
\underset{u_1, u_2, \cdots}{\mathrm{minimize}} \lim_{K\rightarrow +\infty} \frac{\bar{K}}{K}, \label{eq:obj}
\end{equation}
where $\bar{K}$ is the number of blocked requests, $K$ is the total number slots, and the optimization variables are the SBS's actions $\{u_1, u_2, \ldots\}$. We aim to find the optimal policy $\{\mu_1, \mu_2, \ldots\}$, where the policy $\mu_k$ is a mapping from the state space to the action space, i.e., the action in slot $k$ is $u_k = \mu_k(x_k)$, where the system state $x_k$ will be detailed in the next section. According to the statistics of the system including the energy arrival distribution, content request generating rate, content popularity and updating process, we can find the optimal policy via dynamic programming, which will be detailed in the next section.

\section{Optimal Policy with Finite Battery Capacity} \label{sec:MDP}
To find the optimal solution for the problem (\ref{eq:obj}), we need to decide the SBS's action based on the system state at the beginning of each slot. MDP \cite{bertsekas2005dynamic} is an effective mathematical framework to formulate this type of decision making problems. DP algorithm is widely used to deal with the control optimization of stochastic process by breaking it down into a collection of simpler per-stage\footnote{In this paper, the term ``stage" is equivalent to the term ``slot".} subproblems which only depends on the current system state. A standard MDP problem contains the following elements: states, actions, cost function, and state transition. Notice that the energy states and the channel states are continuous in our problem. To make the problem tractable, we first discretize the system states, and then re-formulate our problem as an MDP optimization and use policy iteration algorithm to find the optimal policy.

\subsection{System State} \label{sec:discret}
The system state in stage $k$ is denoted by $x_k = (E_k, Q_k, I_k, C_k),$ where the energy state $E_k$ is the amount of energy in the battery, the request state $Q_k$ is the energy required for the unicast, the push state $C_k$ is the number of pushed contents, and $I_k \in \{0, 1\}$ indicates which content is requested. $I_k = 1$ if the required content is $c_{C_k+1}$, and $I_k = 0$ if the rank of required content is larger than $C_k+1$. The reason to introduce $I_k$ is that if push action is taken, $(C_k+1)$-th ranked content $c_{C_k+1}$ will be transmitted as we always push the most popular content in the list $\widetilde{\bm{C}}_k = (c_{C_k+1}, \ldots, c_N)^T$. Thus the user request for $c_{C_k+1}$ is simultaneously satisfied. Recall that $Q_k$ represents the energy consumption for unicast, which is calculated as $Q_k = E_u(d) = P_t(d)T_p$, where $P_t(d)$ is the transmission power obtained by solving (\ref{eq:rate}). And $Q_k=0$ represents that there is no content request. Denote the state space as $\mathcal{S}$.

To make the DP algorithm tractable, we discretize the energy state $E_k$ and the content request state $Q_k$ into finite sets. Although, the optimal solution for the discretized finite state problem may not be optimal for the continuous state problem, the gap diminishes as the discretization becomes finer {\cite[Vol.~I, Sec.~6.6.1]{bertsekas2005dynamic}}. Specifically, the energy is discretized with unit energy $E_{\mathrm{unit}}$. Then the energy state can be denoted by $E_k \in \{0, 1, \ldots, E_{\mathrm{max}}\}$ with $E_{\mathrm{max}}E_{\mathrm{unit}} = B_{\mathrm{max}}$. $E_k = i$ corresponds to $iE_{\mathrm{unit}}$ amount of energy in the battery, and similarly, $A_k = i$ corresponds to $iE_{\mathrm{unit}}$ amount of energy arrived in stage $k$.

To discretize $Q_k$, we select a series of distances $0<d_1 < d_2 < \ldots < d_M = R$ so that $P_t(d_i)T_p = l_iE_{\mathrm{unit}}$ where $l_i$ is a positive integer for any $i = 1, 2, \ldots, M$. If a user's distance $d$ to the SBS satisfies $d_{i-1} < d \le d_{i}$, we unicast the requested content with energy $P_t(d_i)T_p$, which guarantees the average data rate $r_0$ for all the users in this area. And we set $l_0=0$ denoting that the required energy for unicast is zero. Then the content request state can be denoted by $Q_k \in \{0, 1, \ldots, M\}$, where $Q_k = i$ corresponds to the case that $l_iE_{\mathrm{unit}}$ amount of energy is required for unicast. With the discretization, the state space $\mathcal{S}$ is of dimension $(E_{\mathrm{max}}+1)\times (2M+1) \times (N+1)$.

\subsection{MDP Problem Formulation and Optimization} \label{sec:mdpform}
Besides the system state, we need to further clarify the action, the cost function and the state transition to complete the MDP problem formulation. The action has been modeled as (\ref{eq:uk}). Notice that in different states, the SBS may not be able to take all the three actions. A simple example is that if $E_k = 0$, the SBS can do nothing but sleep, i.e., $u_k = 0$. Hence, the action space is state-dependent, which can be expressed as $u_k \in \mathcal{U}_k(x_k)$. If $E_k \ge l_{Q_k} \ge 1$, i.e., the energy for unicast is available, we have $ 1 \in \mathcal{U}_k(x_k)$. On the other hand, if $E_k \ge E_p$ and $C_k < N$, i.e., the energy for push is available and there are contents to be pushed, we have $ 2 \in \mathcal{U}_k(x_k)$.

The per-stage cost function $g_k(x_k, u_k)$ denotes whether a content request is blocked or not. Mathematically, it can be expressed as
\begin{equation}
g_k(x_k, u_k) = \left\{ \begin{array}{ll} 1, & \textrm{if~} Q_k > 0, u_k \neq 1, \textrm{and~} I_ku_k = 0 \\
0. & \textrm{otherwise}\end{array}\right.
\end{equation}
Notice that a content request is blocked if there is a request but the requested content is not transmitted by either unicast or push. When $g_k(x_k, u_k) = 0$, it refers to either there is no unicast request or the requested content is in the pushed list $\bm{C}_k$.

The state transition is expressed as the conditional probability
\begin{align}
p_{x_k\rightarrow x_{k+1}|u_k} 
= &\mathrm{Pr}(E_{k+1}, Q_{k+1}, I_{k+1}, C_{k+1}|E_k, Q_k, I_{k}, C_k, u_k)\nonumber\\
= &\mathrm{Pr}(\!E_{k+1}|E_k, \!Q_k, \!u_k)\mathrm{Pr}(\!C_{k+1}|C_k,\! u_k) \mathrm{Pr}(\!Q_{k+1}, \! I_{k+1}|C_{k+1}), \label{eq:statetrans}
\end{align}
where the second equality is derived by the definition of conditional probability. We calculate the state transition probability according to (\ref{eq:statetrans}). Firstly, we denote $p_a(i), i = 0, 1, \ldots$ as the probability that $iE_{\mathrm{unit}}$ amount of energy arrives, which satisfies $p_a(i) \in [0, 1]$ and $\sum_i p_a(i) = 1$. To simplify the description, we set $p_a(i) = 0, \forall i = -1, -2, \ldots$. Then we have
\begin{align}
\mathrm{Pr}(E_{k+1}|E_k, Q_k, u_k) =
&\left\{ \begin{array}{ll}
p_a(E_{k+1} - E_k),  & \textrm{if~}\! u_k \!=\! 0, \!E_{k+1}\! <\! E_{\mathrm{max}}\\
1-\sum\limits_{i=0}^{E_{\mathrm{max}}-E_k-1}p_a(i), & \textrm{if~}\! u_k \!=\! 0, \!E_{k+1} \!= \! E_{\mathrm{max}}\\
p_a(E_{k+1}-E_k+l_{Q_k}), & \textrm{if~} \!u_k  \!=\! 1, E_{k+1} \!<\! E_{\mathrm{max}}\\
1-\sum\limits_{i=0}^{E_{\mathrm{max}}\!-\!E_k\!+\!l_{Q_k}\!-\!1}p_a(i), & \textrm{if~} \!u_k  \!=\! 1, \!E_{k+1} \!= \! E_{\mathrm{max}}\\
p_a(E_{k+1}-E_k+l_{M}), & \textrm{if~} \!u_k  \!=\! 2,\! E_{k+1} \!<\! E_{\mathrm{max}}\\
1-\sum\limits_{i=0}^{E_{\mathrm{max}}\!-\!E_k\!+\!l_{M}\!-\!1}p_a(i). & \textrm{if~} \!u_k \!=\! 2,\! E_{k+1} \!=\! E_{\mathrm{max}}\\
\end{array}\right. \label{eq:Ektrans}
\end{align}
Notice that when $E_{k+1} = E_{\mathrm{max}}$, the energy arrival may exceed the battery capacity. So the probability is calculated by summarizing all the possible energy arrival conditions.

Secondly, the push state $C_k$ can only transit to its neighboring values $C_k+1, C_k-1$ or keep constant, i.e., $C_{k+1} \in \{\max\{0, C_k-1\}, C_k, \min\{N, C_k+1\}\}$. The transition probability of push state $C_k$ is expressed as
\begin{align}
\mathrm{Pr}(C_{k+1}|C_k, u_k) =
&\left\{ \begin{array}{ll}
p_c\frac{C_k}{N}, & \textrm{if~} u_k\!<\! 2, C_{k+1} \!=\! C_{k}\!-\!1\ge 0, \textrm{~or~} u_k \!=\! 2, C_{k+1}\! =\! C_{k}\!<\!N\\
1-p_c\frac{C_k}{N}, & \textrm{if~} u_k\! <\! 2, C_{k+1} \!=\! C_{k}, \textrm{~or~} u_k \!=\! 2, C_{k+1} \!=\! C_{k}\! +\! 1 \!\le \!N\\
0. & \textrm{else}
\end{array}\right. \label{eq:Cktrans}
\end{align}

Finally, the content request state transition is
\begin{align}
\mathrm{Pr}(Q_{k+1}, I_{k+1}|C_{k+1}) =
&\left\{ \begin{array}{ll}
(1-p_u) + p_u\sum_{i=1}^{C_{k+1}}f_{i}, & \textrm{if~} Q_{k+1} \!=\! 0\\
p_uf_{C_{k+1}+1}\frac{d_{m}^2-d_{m-1}^2}{R^2}, & \textrm{if~} Q_{k+1} \!=\! m \!>\! 0, \textrm{~and~} I_{k+1} \!=\! 1\\
p_u(1-\sum_{i=1}^{C_{k+1}+1}f_{i})\frac{d_{m}^2-d_{m-1}^2}{R^2}, & \textrm{if~} Q_{k+1} \!=\! m \!>\! 0, \textrm{~and~} I_{k+1} \!=\! 0
\end{array}\right. \label{eq:Qktrans}
\end{align}
where $f_i$ is calculated according to (\ref{eq:pop}) and $d_0=0$. The transition probability (\ref{eq:Qktrans}) is calculated based on our model, i.e., there is a content request with probability $p_u$, requesting which content depends on the popularity distribution $f_i$, and the location of the request follows uniform distribution in the cell coverage. Notice that if $C_{k+1} = N$, we always have $Q_{k+1}=0$. In summary, the state transition probability (\ref{eq:statetrans}) is calculated based on the widely-used models about energy arrival, content popularity, and user distribution. In fact, our framework is also applicable to some other models.

Based on the MDP framework, the original optimization problem (\ref{eq:obj}) can be re-written as
\begin{equation}
\min \lim_{K \rightarrow +\infty}\frac{1}{K}\mathbb{E}\left[ \sum_{k=0}^{K-1}g(x_k, \mu_k(x_k))\right]. \label{eq:MDPobj}
\end{equation}
The expectation operation is taken over all the random parameters. The optimization is taken over all the possible policies $\{\mu_1, \mu_2, \ldots\}$. It can be proved that for any two states, there is a stationary policy $\bm{\mu} = \{\mu(x)\}_{x \in \mathcal{S}}$ so that one state can be accessed with non-zero probability from the other with finite steps. Consequently, the optimal value is independent of the initial state $x_0$, and there exists an optimal stationary policy $\bm{\mu}^*$ \cite[Vol.~II, Sec~4.2]{bertsekas2005dynamic}.

According to \cite[Vol.~II, Prop.~4.2.1]{bertsekas2005dynamic}, the optimal average cost $\lambda^*$ together with some vector $\bm{h}^* = \{h^*(x)|x \in \mathcal{S}\}$ satisfies the Bellman's equation
\begin{equation}
\lambda^* + h^*(x) = \min_{u\in \mathcal{U}(x)}\left[ g(x, u) + \sum_{y \in \mathcal{S}}p_{x\rightarrow y | u}h^*(y)\right]. \label{eq:bellman}
\end{equation}
Further more, if $u = \mu^*(x)$ attains the minimum value of (\ref{eq:bellman}) for each $x$, the stationary policy $\bm{\mu}^*$ is optimal. Based on the Bellman's equation, instead of the long term average cost minimization, we only need to deal with (\ref{eq:bellman}) which only relates with per-stage cost $g(x, u)$ and state transition probability $p_{x\rightarrow y | u}$. The policy iteration algorithm \cite[Vol.~II, Sec.~4.4]{bertsekas2005dynamic} can effectively solve the problem.

\subsection{Policy Iteration Algorithm}

The policy iteration algorithm starts with any feasible stationary policy, and improves the objective step by step. Suppose in the $j$-th step, we have a stationary policy denoted by $\bm{\mu}^{(j)}$. Based on this policy, we perform \emph{policy evaluation} \cite[Vol.~II, Sec.~4.4]{bertsekas2005dynamic}, i.e., we solve the following linear equations
\begin{equation}
\lambda^{(j)} + h^{(j)}(x) = g(x, \mu^{(j)}(x)) + \sum_{y \in \mathcal{S}}p_{x\rightarrow y | \mu^{(j)}(x)}h^{(j)}(y) \label{eq:linear}
\end{equation}
for $\forall x \in \mathcal{S}$ to get the average cost $\lambda^{(j)}$ and vector $\bm{h}^{(j)}$. Notice that there are $(E_{\mathrm{max}}+1)\times (2M+1) \times (N+1)$ equations but $(E_{\mathrm{max}}+1)\times (2M+1) \times (N+1)+1$ unknown parameters, hence more than one solutions exist, which are different with each other by a constant value for all $h^{(j)}(x)$. Without loss of generality, we can set for example $h^{(j)}(E_{\mathrm{max}}+1, 2M+1, N_S+1)=0$, then the solution for (\ref{eq:linear}) is unique.

As $\bm{\mu}^{(j)}$ may not be the optimal policy, we subsequently perform \emph{policy improvement} \cite[Vol.~II, Sec.~4.4]{bertsekas2005dynamic} step to find the policy $\bm{\mu}^{(j+1)}$ which minimizes the right hand side of Bellman's equation
\begin{equation}
\mu^{(j+1)}(x) \!=\! \arg\!\min_{u\in \mathcal{U}(x)} \!\left[g(x, u) \!+\! \sum_{y \in \mathcal{S}}\!p_{x\rightarrow y | u}h^{(j)}(y)\right]. \label{eq:ukplus1}
\end{equation}

If $\bm{\mu^{(j+1)}} = \bm{\mu^{(j)}}$, the algorithm terminates, and the optimal policy is obtained $\bm{\mu^*} = \bm{\mu^{(j)}}$. Otherwise, repeat the procedure by replacing $\bm{\mu^{(j)}}$ with $\bm{\mu^{(j+1)}}$. It is proved that the policy iteration algorithm terminates in finite number of iterations \cite[Vol.~II, Prop.~4.4.1]{bertsekas2005dynamic}. 

%
%
%
%
%
%
%
%
%


Although the DP-based solution can obtain the optimal policy, it has two drawbacks. Firstly, it usually encounters the \emph{curse of dimensionality} \cite{bertsekas2005dynamic} when the number of states is quite large. Secondly, it is difficult to obtain the closed-form expressions to reveal the structure of the optimal policy. The algorithm need to be operated for every set of parameters to get the numerical result. To tackle these problems, we try to design some heuristic algorithms in the next section. Notice that still, studying MDP formulation is meaningful as its optimality provides a theoretical upper-bound to evaluate the performance of the other sub-optimal algorithms. In addition, the optimal policy can be efficiently solved by the DP algorithm if the number of system states is relatively small.

\section{{Threshold-based Policies}} \label{sec:thrpolicy}
In this section, we consider some threshold-based policies. The intuition is that pushing the popular contents enhances the energy efficiency as lots of duplicated unicasts can be avoided. However, the push action itself consumes energy. As we always push the most popular contents, there must be a threshold on $C_k$, denoted by $C_{\mathrm{thr}}$, so that the energy consumption for pushing contents in the list $\bm{C} = (c_1, \ldots, c_{C_{\mathrm{thr}}})$ is less than the total energy consumption for duplicated unicasts without push. Also, threshold-based policies are analytically tractable and have some good asymptotic properties, as shown later. Thus, we consider that the SBS always takes push action as long as the number of pushed contents is smaller than the threshold $C_{\mathrm{thr}}$. Notice that as the system is energy limited, the threshold should depend on the energy state $E_k$ as well as the content request state $Q_k$. However, it is difficult to analyze the state-dependent thresholds. Hence, we focus on the policies with constant threshold to obtain some closed-form results. 

There are two steps to design a {threshold-based policy}. Firstly, determine the threshold $C_{\mathrm{thr}}$. Secondly, determine the action that is taken when the threshold $C_{\mathrm{thr}}$ is achieved. With the objective of minimizing the number of requests blocked by the SBS, we first analyze the performance of push action with sufficient energy, which is detailed in the following lemmas. {Notice that our analysis in this section is done in the continuous regime.}

\begin{lemma} \label{lemma:Ppush}
If the required energy for push can always be satisfied, the stationary probability with which the SBS pushes a content in threshold-based policy is
\begin{equation}
\mathrm{Pr}(u_k = 2) = \frac{p_cC_{\mathrm{thr}}}{N+p_c}. \label{eq:puk}
\end{equation}
\end{lemma}
\begin{IEEEproof}
See Appendix \ref{proof:Ppush}.
\end{IEEEproof}

The blocking probability of a threshold-based policy is composed of two parts. The first part is due to push action itself. If the SBS is pushing a content while a user is requesting another content, the request is blocked. The second part is due to energy shortage, i.e., the user request will be blocked if there is not enough energy for unicast. Based on the observation, we can derive the performance lower bound.

\def\bydefn{\stackrel{def}{=}}

\begin{lemma} \label{lemma:LB}
The blocking probability of the threshold-based policy is lower bounded by
\begin{equation}
p_{\mathrm{blk, LB}}(C_{\mathrm{thr}}) =  \frac{p_c p_u C_{\mathrm{thr}}}{N+p_c} (1-\sum_{i=1}^{C_{\mathrm{thr}}}f_i). \label{eq:LB}
\end{equation}
\end{lemma}
\begin{IEEEproof}
See Appendix \ref{proof:LB}.
\end{IEEEproof}

The lower bound (\ref{eq:LB}) is achievable when the required energy for unicast can always be satisfied. While on the other hand, there are some other policies to achieve zero blocking probability when the power supply is sufficient. For example, the SBS can always take unicast action. In this case, the threshold-based policies may not be optimal. However, with insufficient power supply, threshold-based policies can still provide substantial benefits. We will show the performance later by simulations. In the rest of this section, we will analyze several threshold-based policies proposed from different points of view.

\subsection{Push-Only Threshold-based Policy}
Notice that the more the contents are pushed to the user side, the fewer the unicast requests are triggered. Consequently, the number of blocking events due to energy shortage is reduced. Based on this, we consider the POTB policy in which all the harvested energy is used to push contents. If the threshold is achieved, the SBS does not take unicast action but turns to sleep to store energy. According to Lemma \ref{lemma:Ppush}, the threshold for the POTB policy can be chosen according to
\begin{equation}
\frac{p_cC_{\mathrm{thr,PO}}}{N+p_c}E_p \le \bar{A}, \label{eq:CpoCal}
\end{equation}
which means that the average energy consumption for push does not exceed the available harvested energy. Hence, we have
\begin{equation}
C_{\mathrm{thr,PO}} = \min\left\{N, \left\lfloor \frac{(N+p_c)\bar{A}} {p_cE_p} \right\rfloor \right\}, \label{eq:Cpo}
\end{equation}
where $\lfloor x \rfloor$ is the largest integer smaller than $x$. The performance of the POTB policy can be guaranteed by the following theorem.

\begin{theorem} \label{thm:pushonly}
Given the average energy arrival rate $\bar{A}$, if $B_{\mathrm{max}} \rightarrow +\infty$, the blocking probability of POTB policy is
\begin{equation}
p_{\mathrm{blk,PO}} = p_u \Big(1- \sum_{i=1}^{C_{\mathrm{thr,PO}}}f_i\Big), \label{eq:blkPO}
\end{equation}
where $C_{\mathrm{thr,PO}}$ is expressed as (\ref{eq:Cpo}).
\end{theorem}
\begin{IEEEproof}
See Appendix \ref{proof:pushonly}.
\end{IEEEproof}

Notice that (\ref{eq:blkPO}) is equivalent to the blocking probability with average power constraint. It can be considered as an extension of the result in \cite{ozel2012achieving} where the AWGN channel capacity with infinite battery is equal to that with average power constraint. 
When $\bar{A}$ is strictly larger than the required energy to push $C_{\mathrm{thr}}$ contents but is not enough to push $C_{\mathrm{thr}}+1$ contents, the remained energy can be used for unicast. In this way, the performance can be slightly improved. Based on Theorem \ref{thm:pushonly}, if $\bar{A}$ is large enough so that $C_{\mathrm{thr,PO}} = N$, we have $\sum_{i=1}^{C_{\mathrm{thr,PO}}}f_i = 1$ and hence, the blocking probability is 0. Otherwise, $p_{\mathrm{blk,PO}} > p_{\mathrm{blk,LB}}$ holds.

As (\ref{eq:blkPO}) is a decreasing function of $C_{\mathrm{thr,PO}}$, the performance is better with larger threshold. It motivates us to design an improved version of POTB policy, named as \emph{always-push threshold-based (APTB) policy}, which is equivalent to the POTB policy with threshold $C_{\mathrm{thr}} = N$. However, if $\bar{A} < \frac{p_cN}{N+p_c}E_p$, there is not closed-form blocking probability expression for APTB policy as it depends on the distribution of energy arrival process.

\subsection{Energy-Efficient Threshold-based Policy}
There is a tradeoff between the energy consumed for content push and the energy saved by the reduction of duplicated unicasts. From energy-efficiency point of view, the push action should be taken if the energy saved exceeds the energy consumed. The following proposition describes the condition in which content push is the most energy-efficient.

\begin{lemma} \label{lemma:pushcond}
If the index of content $c_i$ satisfies
\begin{equation}
i \le \left( \frac{Np_u\bar{E}_u}{p_c {E}_p\sum_{j=1}^{N}1/j^v}\right)^{\frac{1}{v}}, \label{eq:pushcond}
\end{equation}
where $\bar{E}_u = \int_{0}^{R}E_u(d)\frac{2d}{R^2}\mathrm{d}d$ is the expected energy consumption of unicast, the energy consumption to push the content $c_i$ is no more than the expected energy consumption for unicasting $c_i$ upon requests.
\end{lemma}
\begin{IEEEproof}
See Appendix \ref{proof:pushcond}.
\end{IEEEproof}

Lemma \ref{lemma:pushcond} tells us that if the popularity rank of a content satisfies (\ref{eq:pushcond}), it is more energy efficient to push it to the user side in advance than to unicast it when required. Denote the threshold by
\begin{equation}
C_{\mathrm{thr,EE}} = \min \left\{N, \left\lfloor \left( \frac{Np_u\bar{E}_u}{p_c{E}_p\sum_{j=1}^{N}1/j^v}\right)^{\frac{1}{v}} \right\rfloor \right\}. \label{eq:Cee}
\end{equation}
Then pushing all the contents with rank $i \le C_{\mathrm{thr,EE}}$ is optimal from the energy-efficiency perspective.

With limited power supply from the renewable energy, it is natural to use the harvested energy in an efficient way. For the push action, it is more energy efficient to just push the contents with popularity rank $i \le C_{\mathrm{thr,EE}}$. And for the unicast action, sending a content to a user closer to the SBS is more energy efficient. Based on the above intuition, the EETB policy works as follows. The SBS firstly guarantees that the contents $c_1, \ldots, c_{C_{\mathrm{thr,EE}}}$ are pushed to the user side, and then responds part of the unicast requests for the contents $c_{C_{\mathrm{thr,EE}}+1}, \ldots, c_N$ according to the users' location. It is detailed in Algorithm \ref{alg:EE}.
In this algorithm, (\ref{eq:EEd}) is derived from the condition that the average energy consumption cannot exceed the average energy arrival, i.e.,
\begin{equation}
\frac{p_cC_{\mathrm{thr,EE}}E_p}{N+p_c} + \eta\int_{0}^{\tilde{d}}E_u(d) \frac{2d}{R^2}\mathrm{d}d \le \bar{A},
\end{equation}
where
\begin{equation}
\eta = \Big(1-\frac{p_cC_{\mathrm{thr,EE}}}{N+p_c} \Big) p_u \Big(1- \sum_{i=1}^{C_{\mathrm{thr,EE}}}f_i\Big)
\end{equation}
can be viewed as the probability of generating a unicast request for a content $c_i$, where $i \in \{{C_{\mathrm{thr,EE}}+1}, \ldots, N\}$, and the $\max$ and $\min$ operations in (\ref{eq:EEd}) are used so that if $\bar{A} \le \frac{p_cC_{\mathrm{thr,EE}}E_p}{N+p_c}$ or $\bar{A} \ge \frac{p_cC_{\mathrm{thr,EE}}E_p}{N+p_c} + \eta\bar{E}_u$, the equation still works. Also notice that we adopt the distance metric to decide whether to unicast or not, which can be replaced by the measured average SINR in real system.

\begin{algorithm}[th]
\caption{EETB policy} \label{alg:EE}
\begin{algorithmic}

\STATE Find the maximum $\tilde{d}$ satisfying
\begin{equation} \label{eq:EEd}
\int_{0}^{\tilde{d}}\!E_u(d) \frac{2d}{R^2}\mathrm{d}d \!\le\! \min\{\bar{E}_u,\! \frac{1}{\eta} \max\{0, \! \bar{A} \!-\! \frac{p_cC_{\mathrm{thr,EE}}E_p}{N+p_c}\}\}
\end{equation}

\IF{$C_k < C_{\mathrm{thr,EE}}$}

\STATE \textbf{if} {$E_k \ge {E}_p$} \textbf{then} Push a content to all the users.

\STATE \textbf{else} Keep sleep to store energy.

\STATE \textbf{end if}

\ELSE

\STATE \textbf{if} {$0 < Q_k \le \min\{E_u(\tilde{d}), E_k\}$} \textbf{then} Unicast the requested content to the user.

\STATE \textbf{else} Keep sleep to store energy.

\STATE \textbf{end if}

\ENDIF

\end{algorithmic}
\end{algorithm}
%
%
%
%
%
%
%
%
%
%
%
%
%
%
%

The blocking probability performance of the EETB policy is summarized as follows.
\begin{theorem} \label{thm:EEblk}
If $\bar{A} \ge \frac{p_cC_{\mathrm{thr,EE}}E_p}{N+p_c}$ and $B_{\mathrm{max}} \rightarrow +\infty$, the blocking probability of EETB policy is
\begin{equation}
p_{\mathrm{blk,EE}} = p_{\mathrm{blk,LB}}(C_{\mathrm{thr,EE}}) + \eta\frac{R^2-\tilde{d}^2}{R^2}, \label{eq:pblkEE}
\end{equation}
where $C_{\mathrm{thr,EE}}$ is expressed as (\ref{eq:Cee}).
\end{theorem}
\begin{IEEEproof}
See Appendix \ref{proof:EEblk}.
\end{IEEEproof}

{The result is similar to Theorem \ref{thm:pushonly}, i.e., the blocking probability with infinite battery is equal to that with average power constraint.} In this policy, the lower bound $p_{\mathrm{blk,LB}}$ can be achieved if $\tilde{d} = R$, i.e., all the unicast requests are satisfied. Notice that if the condition $\bar{A} \ge \frac{p_cC_{\mathrm{thr,EE}}E_p}{N+p_c}$ is not satisfied, the EETB policy degenerates to the APTB policy.

{With the closed-form expressions (\ref{eq:blkPO}) and (\ref{eq:pblkEE}), we can easily find a better policy to minimize the blocking probability, i.e, we adopt the corresponding policy with smaller value calculated based on (\ref{eq:blkPO}) and (\ref{eq:pblkEE}).}


%

\subsection{Greedy Optimal Threshold-based Policy}
In the above discussion, the thresholds of the POTB policy and the EETB policy are found by minimizing the number of unicast requests and by maximizing the energy efficiency, respectively. We can also find the optimal threshold-based policy by greedily searching over all the possible thresholds. For each threshold, we calculate the average energy required to maintain the threshold. Secondly, we use the remained energy for energy-efficient unicast, i.e., determine a radius $d_{\mathrm{thr}}$ and serve the content requests generated inside the circle. The procedure is detailed in Algorithm \ref{alg:GO}.

\begin{algorithm}[th]
\caption{GOTB policy} \label{alg:GO}
\begin{algorithmic}[1]

\STATE Set $C_{\mathrm{thr}} = 0, p_{\mathrm{blk}} = 1, d_{\mathrm{thr}} = 0$.

\FOR{$C=0$ to $C_{\mathrm{thr,PO}}$}

\STATE With threshold $C$, calculate $\tilde{d}$ by (\ref{eq:EEd}), and then calculate $p_{\mathrm{blk},C}$ by (\ref{eq:pblkEE}).

\IF{$p_{\mathrm{blk},C} < p_{\mathrm{blk}}$}

\STATE Set $C_{\mathrm{thr}} = C, p_{\mathrm{blk}} = p_{\mathrm{blk},C}, d_{\mathrm{thr}} = \tilde{d}$.

\ENDIF

\ENDFOR

\end{algorithmic}
\end{algorithm}

When the algorithm terminates, the threshold-based policy with parameters $C_{\mathrm{thr}}, d_{\mathrm{thr}}$ is optimal. Notice that the search range is from $0$ to $C_{\mathrm{thr,PO}}$ since when $C > C_{\mathrm{thr,PO}}$, there is no spare energy for unicast. The GOTB policy can be used to evaluate the performance of the POTB policy and the EETB policy. Also notice that all the proposed threshold-based policies only depend on the statistic information of energy, contents, and user traffic. Hence, they are applicable to some other general models.

\subsection{Analysis with Finite Battery Capacity}
With finite battery capacity, the closed-form blocking probability expression cannot be obtained. We analyze the threshold-based policies using FSMC model. Specifically, for a given stationary policy $\mu(x)$, the state transition $p_{x_k\rightarrow x_{k+1}|\mu(x_k)}$ can be calculated based on (\ref{eq:statetrans}) and the equations thereafter. Then we can get the state transition matrix $\mathbf{P}_{\mu}$. By solving $\pi\mathbf{P}_{\mu} = \pi$ and $\sum_x \pi_x = 1, \pi_x \ge 0$, the stationary distribution of the system states can be obtained. Finally, the blocking probability can be calculated as $p_{\mathrm{blk}} = \sum_{x\in \{x: Q>0, u \neq 1 \textrm{~and~} Iu=0\}} \pi_x$.

Notice that for the EETB policy, $\tilde{d}$ in (\ref{eq:EEd}) cannot be arbitrary real value due to the discretization on the content request state. In stead, we have $\tilde{d} \in \{0, d_1, \ldots, d_M\}$. As a result, there might be a gap between the available energy and the energy required for unicast, which causes some quantization error.

\section{Numerical Results} \label{simulation}

We run some simulations to evaluate the performance of proposed algorithms. We set the cell radius $R=50$m, the required content delivery spectrum efficiency $r_0/W = 1$bps/Hz, the pathloss parameters $\beta = 10$dB and $\alpha = 2$. The maximum transmit power or equivalently the transmit power for cell-edge user is set $P_t(R) = 1$Watt. The channel coefficient $h$ follows Rayleigh fading, whose mean value and the interference plus noise power $\sigma^2 + I$ are set so that (\ref{eq:rate}) holds for $r=r_0, d = R, P_t = P_t(R)$. The users are sorted into $M=5$ classes and the energy is quantized with unit $E_{\mathrm{unit}} = \frac{P_t(R)T_p}{M}$. Hence, the number of energy units for push is $E_p = M$, and the content request in class $m$ consumes $m$ units of energy. Assume the energy arrival process follows a Poisson distribution. {Notice that our analytical results do not depend on the energy arrival model, and hence can be applied to some more realistic models.} We set $N = 20$, and the Zipf parameter $v = 1$.

\begin{figure}
\centering
\includegraphics[width=3.4in]{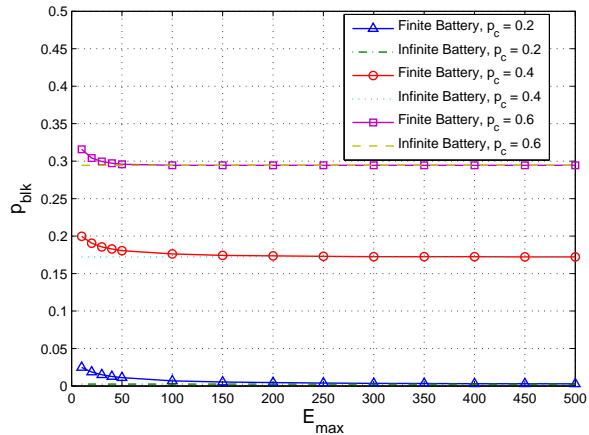}
\caption{The influence of battery capacity on the POTB policy. $\bar{A} = 1.0, p_u = 0.9$.} \label{fig:POvsBmax}
\end{figure}

\begin{figure}
\centering
\includegraphics[width=3.4in]{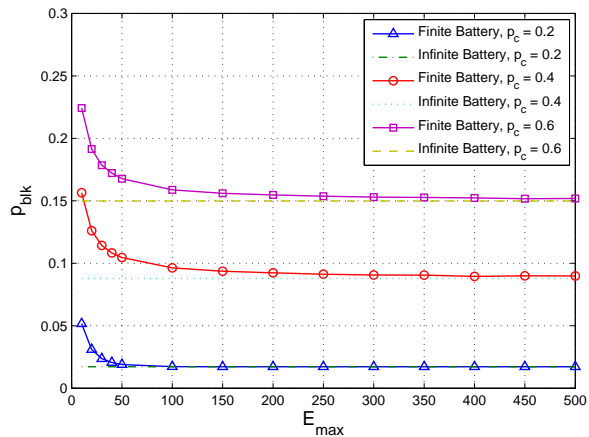}
\caption{The influence of battery capacity on the EETB policy. $\bar{A} = 1.0, p_u = 0.9$.} \label{fig:EEvsBmax}
\end{figure}

Firstly, we set $\bar{A} = 1.0, p_u = 0.9$ in the simulations and evaluate the influence of the battery capacity. The reuslts are depicted in Figs.~\ref{fig:POvsBmax} and \ref{fig:EEvsBmax}. Comparing the two figures, it can be found that the POTB policy is more tolerable to the limited battery capacity than the EETB policy. On the one hand, the performance degradation of the POTB policy w.r.t. the infinite battery case is smaller than that of the EETB policy. For instance, with $p_c = 0.4$ and $E_{\mathrm{max}} = 20$, the blocking probability of the POTB policy increases by 10\%, while the EETB policy increases by more than 40\%. On the other hand, the POTB policy converges to the infinite battery case much faster than the EETB policy. Take $p_c = 0.6$ as an example, the convergence point of the POTB policy is $E_{\mathrm{max}} \approx 50$, but that of the EETB policy is larger than 500. The reason is that if we view the battery as an energy queue, the service process of the POTB policy (only push) is less dynamic than that of the EETB policy (including both push and unicast with variable energy requirements).

Then we evaluate the performance of the POTB policy, the EETB policy as well as the APTB policy by comparing with the GOTB policy, the DP optimal policy and the no-push baseline policy which is termed as \emph{service-on-demand} policy, i.e., contents are not pushed and the users are served upon requests.

\begin{figure}
\centering
\includegraphics[width=3.4in]{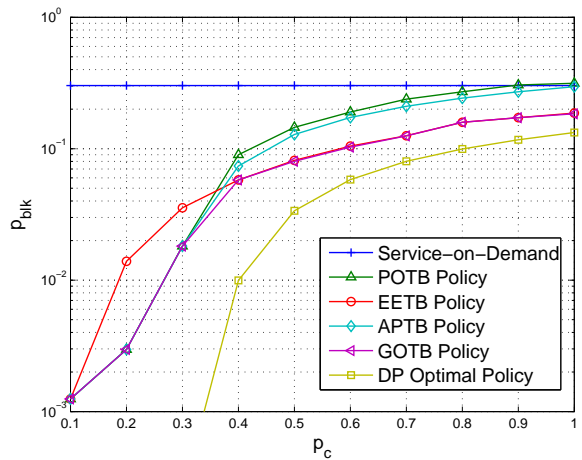}
\caption{The performance comparison versus the content updating rate $p_c$ {with $N=20$}. The energy arrival rate is $\bar{A} = 1.5$ and the content request generating rate is $p_u = 0.9$.} \label{fig:vsPc}
\end{figure}

Fig.~\ref{fig:vsPc} shows the performance comparison by varying the content updating rate $p_c$. We set the energy arrival rate $\bar{A} = 1.5$ and the content request generating rate $p_u = 0.9$. Compared with the service-on-demand policy, the push policies perform better and the gain is enhanced as the content updating rate $p_c$ decreases, which indicates that the benefit of push is more noticeable if the content set is more stable. On the other hand, there is a cross point between the EETB policy and the POTB policy. The EETB policy achieves the same performance with the GOTB policy when $p_c \ge 0.4$, while the POTB policy performs the same when $p_c \le 0.3$. While they converge to the same when $p_c \le 0.1$ since $C_{\mathrm{thr, PO}} = C_{\mathrm{thr, EE}} = N$ in this situation. Also, the APTB policy shows a slight performance gain compared with the POTB policy. However, there is still gap between the threshold-based policies and the DP optimal policy.

\begin{figure}
\centering
\includegraphics[width=3.4in]{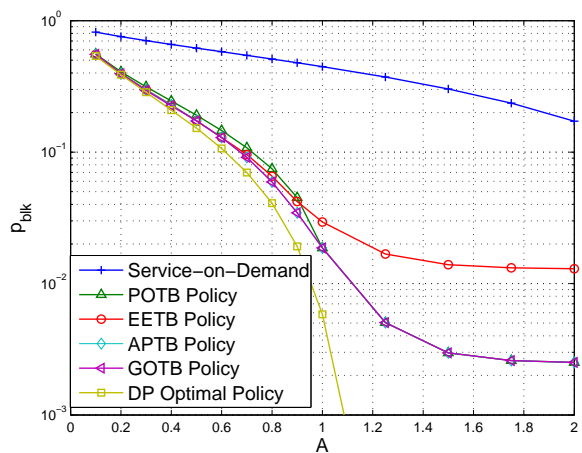}
\caption{The performance comparison versus the energy arrival rate $\bar{A} = A$. The content updating rate is $p_c = 0.2$, and the content request generating rate is $p_u = 0.9$.} \label{fig:vsE}
\end{figure}

We further compare the policies by varying the energy arrival rate $\bar{A} = A$ with $p_c = 0.2$ and $p_u = 0.9$. The result is illustrated in Fig.~\ref{fig:vsE}. As the content updating rate is relatively small, the performance gain of push-based policies is remarkable compared with the non-push policy. The APTB policy performs the same with the GOTB policy, while the POTB policy and the EETB policy perform the same as the GOTB policy in high energy arrival rate regime and low energy arrival rate regime, respectively. Notice that the curves corresponding to the threshold-based policies become flat as $\bar{A}$ is large enough, which illustrates the performance lower bound. Finally, compared with the DP optimal policy, the performance gap between it and the threshold-based policies diminishes when $\bar{A}$ becomes small.

\begin{figure}
\centering
\includegraphics[width=3.4in]{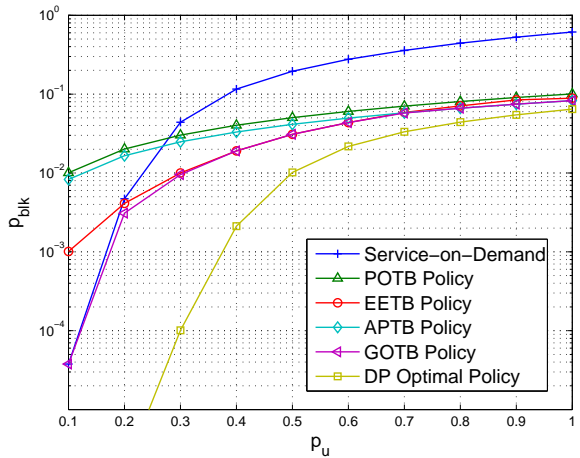}
\caption{The performance comparison versus the content request generating rate $p_u$. The energy arrival rate is $\bar{A} = 0.75$, and the content updating rate is $p_c = 0.2$.} \label{fig:vsPu}
\end{figure}

At last, the performance variation w.r.t. the content request generating rate $p_u$ is depicted in Fig.~\ref{fig:vsPu}. Notice that except the GOTB policy, our proposed threshold-based policies perform worse than the service-on-demand policy when the content request generating rate is very low. This is because if the content request arrives very slowly (even slower than the content updating rate), each content is requested very few times (no larger than 1 with high probability). It is meaningless to push contents in advance. However, as the search process in the GOTB policy contains the case $C=0$, it will ultimately converge to the service-on-demand policy as $p_u$ decreases. Compared with the GOTB policy, there is a cross point between the EETB policy curve and the APTB policy curve. The former performs the same with the GOTB policy with low content request generating rate (lower than 0.7 but not lower than 0.3), the latter does with high content request generating rate (higher than 0.7). In addition the performance gap between the threshold-based policies and the DP optimal policy becomes small as $p_u$ increases.

\section{Conclusion} \label{concl}
In this paper, proactive push policies in EH based SBS were studied. With infinite battery capacity, two threshold-based policies, the POTB policy and the EETB policy, achieve the closed-form blocking probability the same with average power constraint. With finite battery capacity, we compared the threshold-based policies with the optimal policy and the conventional service-on-demand policy. From the numerical simulations, we found that tremendous performance gain can be obtained through push when the content updating rate is low or the {content request generating rate} is high. We also find that there are cross points between the POTB policy and the EETB policy. The POTB policy performs better than the EETB policy if the energy arrival rate is high or the content updating rate is low. In addition, the POTB policy is more tolerable to the limited battery capacity than the EETB policy. Besides, the threshold-based policies perform close to the optimal policy in the following cases: the energy arrival rate is low, or the {content request generating rate} is high. Future work can consider communication overheads of push mechanism, such as the content fetching and caching cost at the SBS, the content storage cost at the users, and etc. Besides, inter-SBS cooperative content caching and push in multiple SBSs scenario will also be an interesting future research direction.

\appendices

\section{Proof of Lemma \ref{lemma:Ppush}} \label{proof:Ppush}
If the energy for push is always satisfied, $C_k$ will not be smaller than $C_{\mathrm{thr}}-1$, since every time that $C_k = C_{\mathrm{thr}}-1$, a new content is pushed ($u_k = 2$). Thus, there are in total two stationary push states $C_k \in \{ C_{\mathrm{thr}}, C_{\mathrm{thr}}-1\}$, and the push action is always taken in state $C_k = C_{\mathrm{thr}}-1$. We have the transition probabilities
$
\mathrm{Pr}(C_{k+1}=C_{\mathrm{thr}}-1 | C_{k}=C_{\mathrm{thr}}) = p_c\frac{C_{\mathrm{thr}}}{N},
\mathrm{Pr}(C_{k+1}=C_{\mathrm{thr}}-1 | C_{k}=C_{\mathrm{thr}}-1) = p_c\frac{C_{\mathrm{thr}}-1}{N}.
$
As a result, the push state transition matrix can be expressed as
\begin{align}
\mathbf{P_C} = \left( \begin{array}{ll} p_c\frac{C_{\mathrm{thr}}-1}{N} & 1-p_c\frac{C_{\mathrm{thr}}-1}{N}\\
p_c\frac{C_{\mathrm{thr}}}{N} & 1-p_c\frac{C_{\mathrm{thr}}}{N} \end{array}\right).
\end{align}

By solving the equilibrium equation $\bm{\pi} = \bm{\pi}\mathbf{P_C}$, {where $\bm{\pi} = (\pi_{C_{\mathrm{thr}}-1}, \pi_{C_{\mathrm{thr}}}), \pi_i = \mathrm{Pr}(C_k=i), i = C_{\mathrm{thr}}-1, C_{\mathrm{thr}}$ are the stationary probabilities of push state}, we can get
\begin{equation}
\mathrm{Pr}(u_k=2) = \pi_{C_{\mathrm{thr}}-1} = \frac{p_cC_{\mathrm{thr}}}{N+p_c}.
\end{equation}

\section{Proof of Lemma \ref{lemma:LB}} \label{proof:LB}
We derive the lower bound by assuming that there is sufficient power supply for any actions. Thus, a request is blocked only when the SBS is pushing a content while simultaneously another content is requested. Depending on (\ref{eq:Qktrans}) and (\ref{eq:puk}), the {stationary} blocking probability satisfies
\begin{align}
p_{\mathrm{blk}} &\ge \mathrm{Pr}(Q_k > 0, I_k = 0, u_k=2)\nonumber\\
&= \mathrm{Pr}(u_k = 2) \mathrm{Pr}(Q_k > 0, I_k = 0|u_k = 2)\nonumber\\
&= \frac{p_cC_{\mathrm{thr}}}{N+p_c} p_u(1-\sum_{i=1}^{C_{\mathrm{thr}}}f_i) \bydefn p_{\mathrm{blk, LB}}(C_{\mathrm{thr}}).
\end{align}

\section{Proof of Theorem \ref{thm:pushonly}} \label{proof:pushonly}
{For simplicity, we denote $C = C_{\mathrm{thr,PO}}$. Firstly, we assume that there is sufficient power supply for push. In this case, the request for a content in the remained list $\widetilde{\bm{C}}_k = (c_{C+1}, \ldots, c_N)$ will be blocked. Obviously, the stationary blocking probability is expressed as (\ref{eq:blkPO}).

Next, we prove that the blocking probability with energy harvesting and infinite battery capacity is also (\ref{eq:blkPO}). It is equivalent to show that the stationary energy shortage probability, denoted by $p_s$, is zero. We prove it by contradiction. Specifically, we assume that
\begin{align}
p_s = \mathrm{Pr}(E_k < E_p|C_k = C-1) > 0 \label{eq:assume}
\end{align}
and find the contradiction.

We consider the content updating process and the energy process separately. For the content updating process, we introduce the concept of \emph{push request indicator} denoted by $D_k$. When a pushed content leaves, a new content needs to be pushed, i.e., a push request is generated. We use $D_k = 1$ to indicate that a push request is generated, and $D_k = 0$ otherwise. The push state achieves $C$ only if all the push requests are satisfied. By definition, when a push request is generated, it is equivalent to that a pushed content leaves. Therefore, a push request will be generated with probability $\mathrm{Pr}(D_{k+1} = 1|C_k = i) = \frac{p_ci}{N}, i \le C$.

Denote the stationary distribution of $C_k$ by $\pi_i = \mathrm{Pr}(C_k = i), i = 0, \ldots, C$, which satisfies $\pi_i \ge 0$ and $\sum_{i=0}^{C}\pi_i =1$. By the law of total probability, the stationary probability of push request can be calculated as
\begin{align}
\mathrm{Pr}(D_{k} = 1) =& \sum_{i = 0}^C \mathrm{Pr}(C_{k-1} = i)\mathrm{Pr}(D_{k} = 1 | C_{k-1} = i) = \sum_{i = 0}^C\pi_{i}\frac{p_ci}{N} \nonumber\\
\le& \pi_C\frac{p_cC}{N} + \sum_{i = 0}^{C-1}\pi_{i}\frac{p_c(C-1)}{N} = \pi_C\frac{p_c}{N}+ \frac{p_c(C-1)}{N}, \label{eq:gamma}
\end{align}
where $\pi_C$ can be calculated by solving the equilibrium equation $\bm{\pi} = \bm{\pi}\mathbf{P_C}$, where $\bm{\pi} = (\pi_0, \ldots, \pi_C)$, and the elements (denoted by $p_{i,j} = \mathrm{Pr}(C_{k+1} = j|C_{k} = i), 0 \le i \le C, 0 \le j \le C$) of the state transition probability matrix $\mathbf{P_C}$ satisfy that: (1) $\sum_j p_{i,j} = 1, p_{i,j} \ge 0, \forall i$; (2) if $|i-j| > 1$, $p_{i,j} = 0$. The expression of $\pi_C$ is
\begin{align}
\pi_C = & \bigg( 1+ \frac{p_{C,C-1}}{p_{C-1,C}} + \cdots + \frac{\prod_{n=1}^C p_{n,n-1}}{\prod_{n=1}^C p_{n-1,n}} \bigg)^{-1}. \label{eq:piC}
\end{align}

We compare with sufficient energy input case as in Appendix \ref{proof:Ppush}, in which only two stationary states $C_k \in \{C, C-1\}$ are accessible, and the push request generating probability can be calculated similarly
\begin{align}
\mathrm{Pr}'(D_{k} = 1) = \pi_C'\frac{p_c}{N}+ \frac{p_c(C-1)}{N}, \label{eq:gammap}
\end{align}
where the stationary distribution
\begin{align}
\pi_C' = & \bigg( 1+ \frac{p_{C,C-1}'}{p_{C-1,C}'} \bigg)^{-1} \label{eq:piCp}
\end{align}
is obtained by solving $\bm{\pi}' = \bm{\pi}'\mathbf{P_C}'$ where the elements of transition probability matrix $\mathbf{P_C}'$ satisfies $p_{i,j}' = 0, \forall i < C-1, j < C-1$. As
\begin{align}
p_{C,C-1}' = p_{C,C-1} = p_c\frac{C}{N},
\end{align}
based on our assumption $p_s > 0$, we have
\begin{align}
p_{C-1, C} =& p_s\mathrm{Pr}(C_{k+1} \!=\! C|C_k \!=\! C\!-\!1, E_k \!<\! E_p) \!+\!
(1\!-\!p_s)\mathrm{Pr}(C_{k+1} \!=\! C|C_k \!=\! C\!-\!1, E_k \!>\! E_p) \nonumber\\
=& p_s\cdot 0 + (1-p_s)p_{C-1,C}' < p_{C-1,C}'.
\end{align}
Comparing (\ref{eq:piC}) and (\ref{eq:piCp}), we have $\pi_C' > \pi_C$, which results in $\mathrm{Pr}(D_{k} = 1) < \mathrm{Pr}'(D_{k} = 1)$ according to (\ref{eq:gamma}) and (\ref{eq:gammap}). On the other hand, we have $\pi_C' = 1-\frac{p_cC}{N+p_c}$ as derived in Appendix \ref{proof:Ppush}. Thus, we have
\begin{align}
\mathrm{Pr}(D_{k} = 1) < \frac{p_cC}{N+p_c}.
\end{align}

For energy process, each push request needs to be satisfied with energy usage of $E_p$. The average energy consumption for push satisfies
\begin{align}
\mathrm{Pr}(D_{k} = 1) E_p < \frac{p_cC}{N+p_c}E_p \le \bar{A}. \label{eq:energyq}
\end{align}
{If we consider the energy buffer as a queuing system where the energy units are viewed as customers, (\ref{eq:energyq}) tells that the customer (energy) arrival rate $\bar{A}$ is strictly larger than the service (energy consumption) rate. In this case, the queuing system is unstable, i.e., the arrived customers cannot all be served and the queue overflows \cite[pp.~4]{kleinrock1975queueing}.} As a result, the energy shortage probability is zero, i.e., $p_s = 0$, which contradicts with the assumption (\ref{eq:assume}). Hence, the theorem is proved.}


\section{Proof of Lemma \ref{lemma:pushcond}} \label{proof:pushcond}
Denote $N_p$ as the number of slots in which a content stays in the content list $\bm{C}$. According to the content updating process, $N_p$ follows the geometric distribution, i.e.,
\begin{equation}
\mathrm{Pr}(N_p = n) = (1-p_c\frac{1}{N})^{n-1}p_c\frac{1}{N}.
\end{equation}
Denote $N_r$ as the number of requests for the content $c_i$. In the case that the content $c_i$ stays in the system for $n$ slots, $N_r$ follows the binomial distribution
\begin{equation}
\mathrm{Pr}(N_r = k| N_p = n) = \Big( \begin{array}{l} n \\ k\end{array}\Big)(p_uf_i)^k (1-p_uf_i)^{n-k}.
\end{equation}
By the law of total expectation, the average number of requests for the content with rank $i$ is
\begin{align}
\mathbb{E}[N_r] &= \mathbb{E}_{N_p}[\mathbb{E}[N_r|N_p]]\nonumber\\
&= \sum_{n\ge1}\mathrm{Pr}(N_p = n)np_uf_i \nonumber\\
&= \mathbb{E}[N_p]p_uf_i = \frac{N}{p_c}p_uf_i.
\end{align}

Since the channel state is independent of the content request, the expected unicast energy consumption for all these requests is
\begin{align}
\mathbb{E}[E_u] &= \mathbb{E}[N_r]\bar{E}_u \nonumber\\
&= \frac{N}{p_c}p_uf_i\bar{E}_u. \label{eq:expectEu}
\end{align}
If the content of rank $i$ is pushed to the users, the requests for this content are satisfied by the users' local storage. Hence, the unicast energy consumption (\ref{eq:expectEu}) is avoided. It can be found that pushing the content of rank $i$ can reduce the overall energy consumption if
\begin{equation}
{E}_p \le \frac{N}{p_c}p_uf_i\bar{E}_u.
\end{equation}
With some simple derivation, the condition (\ref{eq:pushcond}) is obtained.

\section{Proof of Theorem \ref{thm:EEblk}} \label{proof:EEblk}
We prove the theorem by contradiction similar to Appendix \ref{proof:pushonly}. Denote $C = C_{\mathrm{thr, EE}}$. If the required energy for EETB policy is always available, a request is blocked when either the SBS is pushing a content, or the user's distance to the SBS is larger than $\tilde{d}$. Thus the {stationary} blocking probability is
\begin{align}
p_{\mathrm{blk}}  = \mathrm{Pr}(Q_k>0, I_k = 0, u_k=2) + \mathrm{Pr}(u_k \neq 2) \mathrm{Pr}(Q_k>E_u(\tilde{d})|u_k \neq 2),
\end{align}
which results in (\ref{eq:pblkEE}). Then we assume the non-zero energy shortage probability and find the contradiction. The difference with Appendix \ref{proof:pushonly} is that the energy shortage may happen in either push state or unicast state. Firstly, we assume the energy is insufficient for push, i.e., {$p_s = \mathrm{Pr}(E_k < E_p|C_k = C-1) > 0$}. Similar with Appendix \ref{proof:pushonly}, we can infer that $\pi_C < \pi_C'$ due to $p_s > 0$, and the required energy satisfies
\begin{align}
&\mathrm{Pr}(D_k = 1) E_p + \pi_Cp_u \Big(1- \sum_{i=1}^{C}f_i\Big)\int_{0}^{\tilde{d}}E_u(d) \frac{2d}{R^2}\mathrm{d}d \nonumber\\
<& \frac{p_cC}{N+p_c} E_p + \pi_C'p_u \Big(1- \sum_{i=1}^{C}f_i\Big)\int_{0}^{\tilde{d}}E_u(d) \frac{2d}{R^2}\mathrm{d}d \le \bar{A}. \label{eq:EpEu1}
\end{align}

Secondly, we assume the energy is insufficient for unicast, i.e., $\int_{0}^{\tilde{d}}p_d(d)\mathrm{d}d > 0$, where $p_d(d) = \mathrm{Pr}(E_k < E_u(d)|Q_k = E_u(d)), d > 0$. Then the required energy satisfies
\begin{equation}
\frac{p_cC}{N+p_c}E_p + \eta\int_{0}^{\tilde{d}}(1-p_d(d))E_u(d) \frac{2d}{R^2}\mathrm{d}d < \bar{A}. \label{eq:EpEu2}
\end{equation}

In both cases, the energy queue overflows. Hence, both (\ref{eq:EpEu1}) and (\ref{eq:EpEu2}) contradict the non-zero energy shortage probability assumption, and the theorem is proved.



\bibliographystyle{IEEEtran}
\bibliography{ref}

\end{document}